\documentclass[11pt]{article}
\usepackage{moriond,epsfig}
\usepackage{epsfig}
\usepackage{fancyhdr}
\usepackage{latexsym}
\usepackage{amsmath}
\usepackage{amssymb}
\usepackage[latin1]{inputenc}
\usepackage{layout}

\def\Journal#1#2#3#4{{#1} {\bf #2}, #3 (#4)}

\def\etal{et al.}

\def\NPB{{\em Nucl. Phys.} {\bf  B}}

\def\PRD{{\em Phys. Rev.} {\bf D}}

\def\EPJ{{\em Eur. Phys. J.} {\bf C}}

\def\MPA{{\em Mod. Phys. Lett.} {\bf A}}
\def\JMA{{\em Int. J. Mod. Phys.} {\bf A}}

\newcommand{\gevc}{\ensuremath{{\rm GeV}\!/c}}

\newcommand{\gevcc}{\ensuremath{{\rm GeV}\!/c^2}}
\newcommand{\tevcc}{\ensuremath{{\rm TeV}\!/c^2}}

\newcommand{\sq}{\ensuremath{{\tilde{\rm q}}}}
\newcommand{\glu}{\ensuremath{{\tilde{\rm g}}}}

\newcommand{\mglu}{\ensuremath{m_\glu}}

\newcommand{\sqqbar}{\ensuremath{\sq\bar{\rm q}}}
\newcommand{\qqbar}{\ensuremath{{\rm q\bar{q}}}}

\newcommand{\chanc}{\ensuremath{{\rm p p \rightarrow \glu \glu }}}
\newcommand{\chand}{\ensuremath{{\rm p p \rightarrow \sq \glu \rightarrow \glu \glu q }}}
\newcommand{\chane}{\ensuremath{{\rm p p \rightarrow \sq \sq \rightarrow \glu \glu q q}}}
\newcommand{\chancc}{\ensuremath{{\rm g g \rightarrow \glu \glu }}}

\def\be{\begin{equation}}
\def\ee{\end{equation}}
\def\bea{\begin{eqnarray}}
\def\eea{\end{eqnarray}}

\begin{document}
\vspace*{4cm}
\title{Interactions of R-hadrons in ATLAS}

\author{ A.C. Kraan }

\address{University of Pennsylvania, Philadelphia, USA}

\maketitle\abstracts{In this talk the detection possibilities of R-hadrons in the ATLAS detector are studied. R-hadrons are stable hadronized gluinos, predicted by certain supersymmetric models. Making use of fully simulated R-hadrons, signatures of single R-hadrons in the ATLAS subdetector are studied, and triggering issues are addressed. Fast simulation is used to study the discovery potential. The abundant production of R-hadrons at the LHC, will allow a quick discovery for a broad range of masses. 
}

\section{Introduction}Heavy stable hadrons are predicted in several models of physics beyond the Standard Model (SM). For example, supersymmetry (SUSY) models exist in which the gluino is stable, including gauge mediated supersymmetry breaking, string motivated models and split supersymmetry~\cite{models}. If R-parity is conserved, a stable gluino would hadronize into heavy (charged and neutral) bound states (for example \glu g, \glu \qqbar, \glu qqq, \sqqbar, \sq qq) called R--hadrons. Predicted masses range from the order of 100 GeV up to a few TeV. Besides SUSY, other extensions of the SM have been proposed predicting new heavy hadrons. Heavy stable particles have been extensively searched for, see e.g. Ref~\cite{pdg}. The phenomenology of stable gluinos has been studied previously~\cite{tilman}. 

\begin{figure}[t]

\begin{picture}(120,110)
\put(0,-10){\epsfxsize45mm\epsfbox{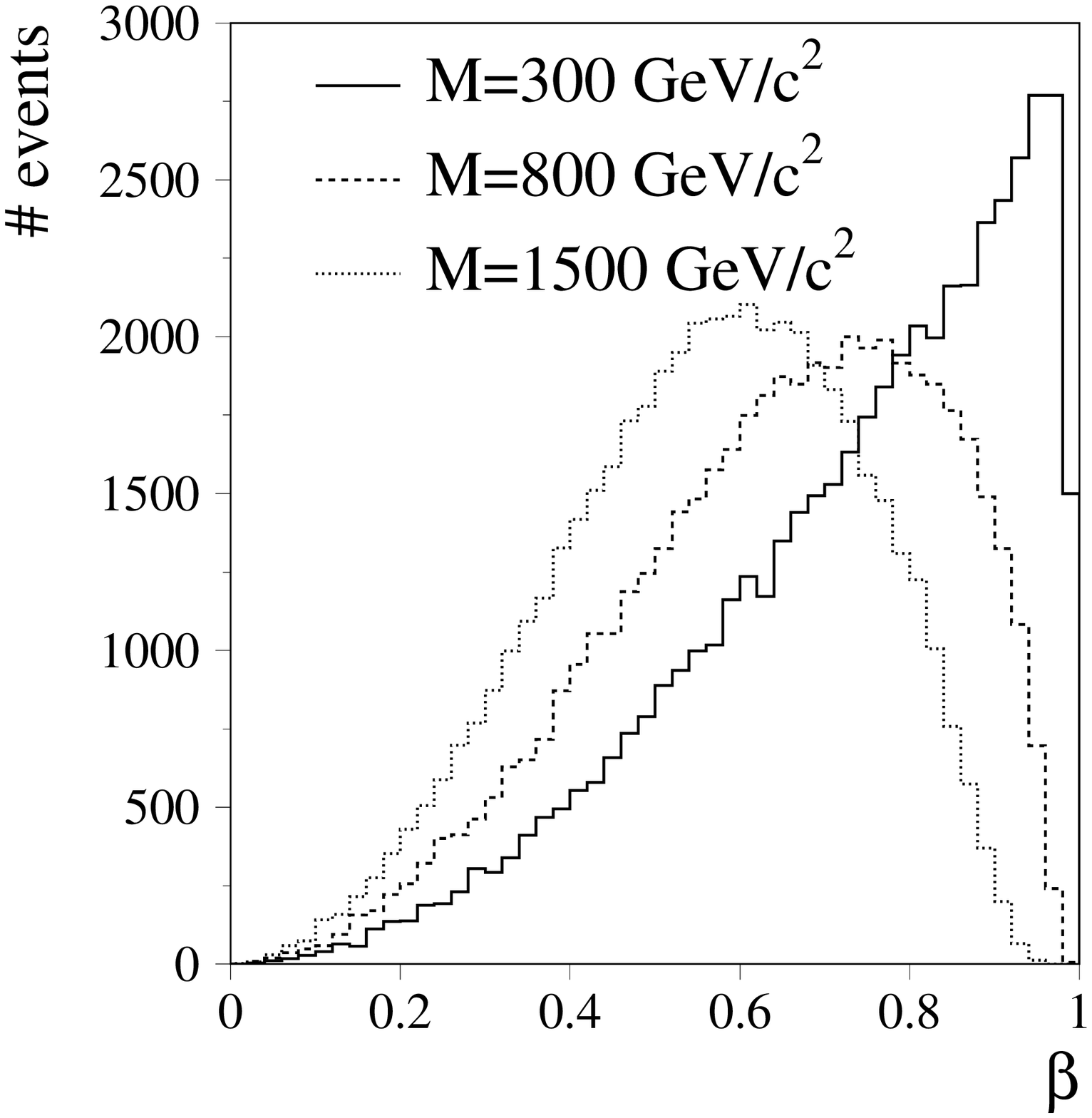}}
\put(40,50){(a)}
\put(150,-10){\epsfxsize45mm\epsfbox{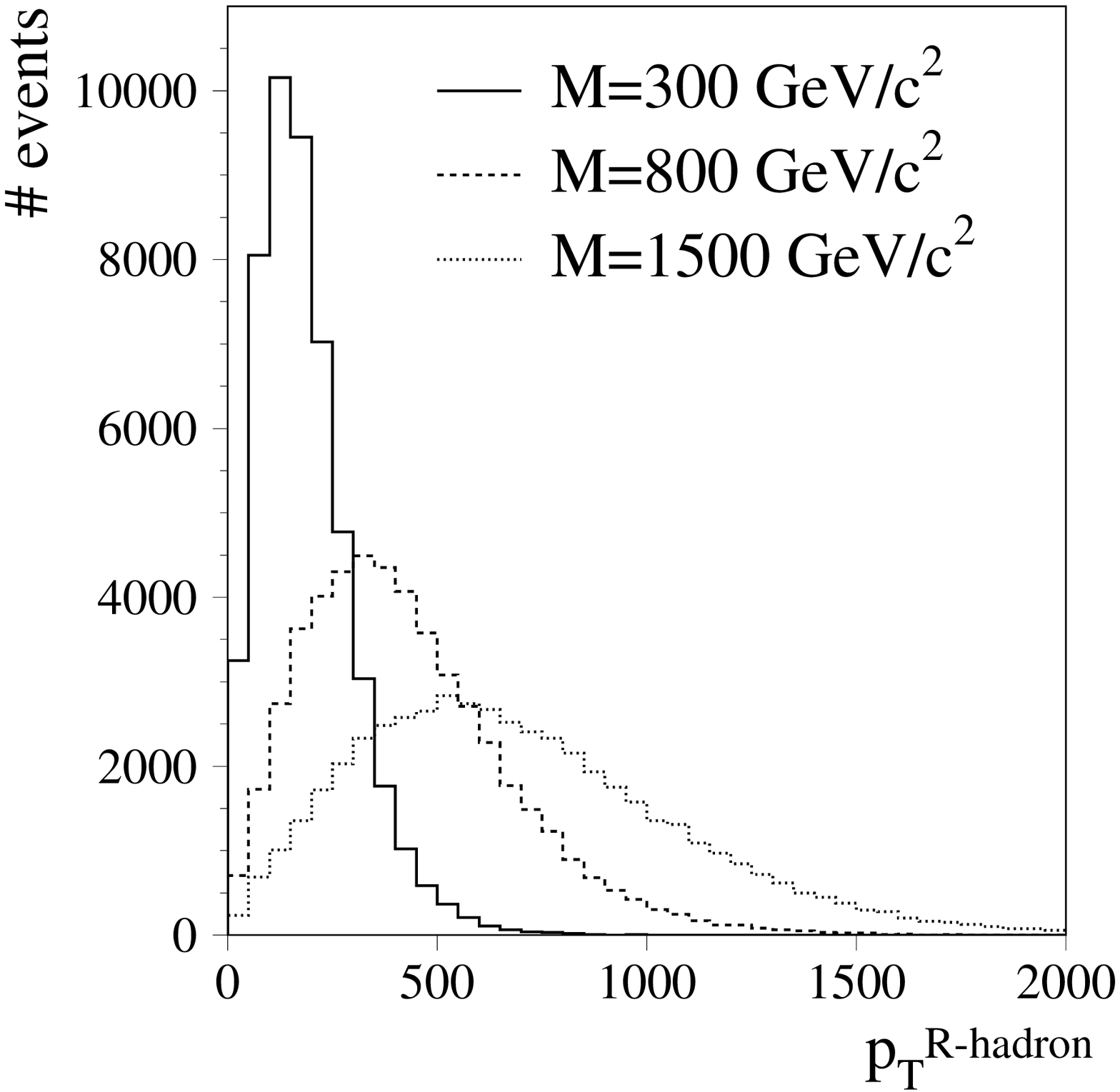}}
\put(240,55){(b)}
\put(300,-10){\epsfxsize45mm\epsfbox{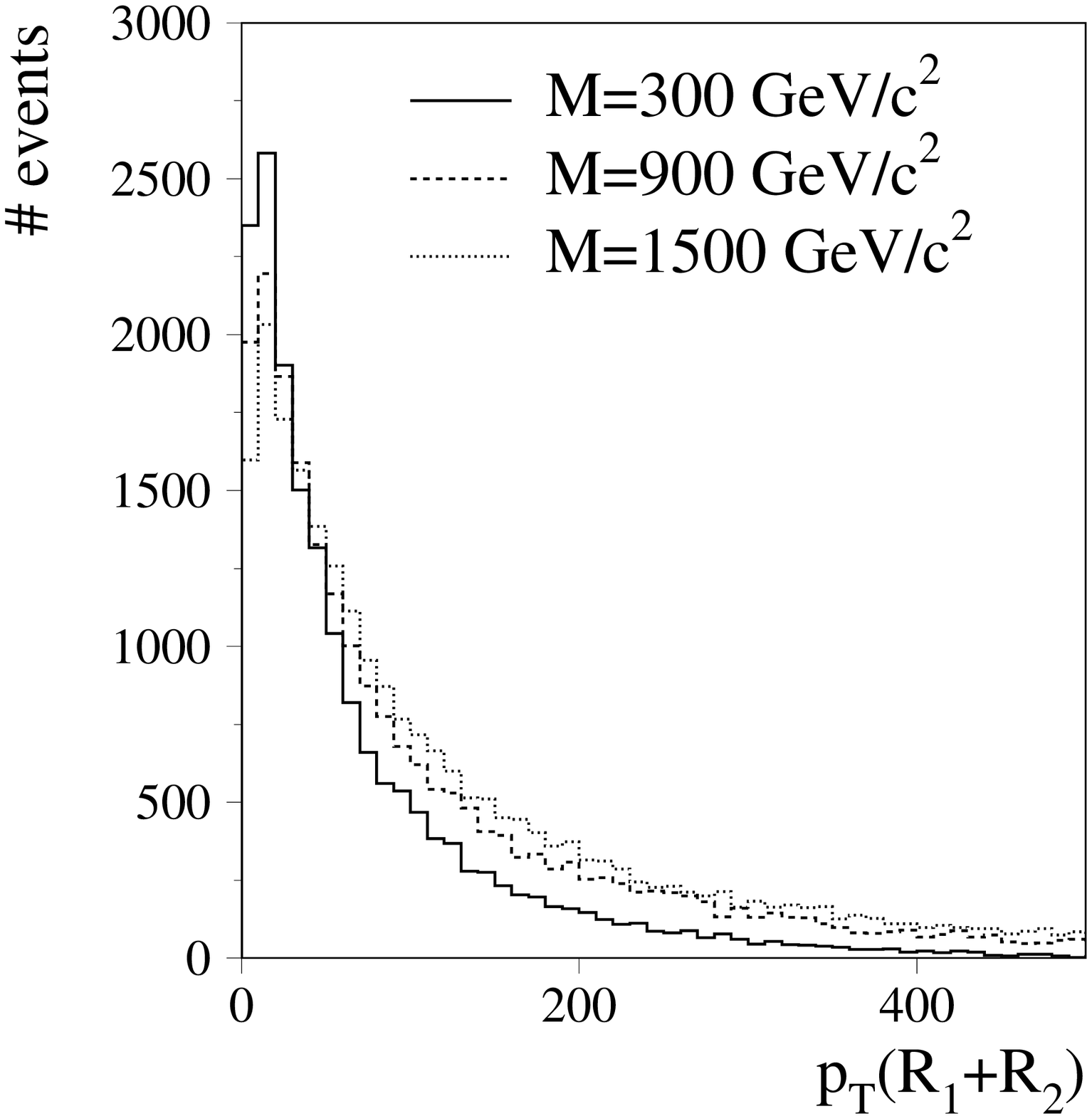}}
\put(390,55){(c)}
\end{picture}
\caption[ ]
{\protect\footnotesize  Aspects of R-hadron production (PYTHIA). (a) The distribution of the velocity of the R-hadrons produced at LHC for \mglu = 300, 800, 1500 \gevcc. Average values are 0.74, 0.64 and 0.56. (b)  The distribution of the transverse momentum of the R-hadrons produced at LHC for \mglu = 300, 800, 1500 \gevcc, with average values 188, 427 and 699 \gevc. (c) The summed transverse momentum of the two R-hadrons, with average values of 74, 129 and 172 \gevc. In the distributions of (a), (b) and (c) a cut for the R-hadrons of $\eta<2.5$ is made.\label{fig:rhadprod}}
\end{figure}

At LHC, gluinos would be directly produced via one of the processes \chanc, \chand, or \chane, where the latter two channels are only accessible if the squark is not too heavy. In this study, we focus on \chancc, because it is purely strong and, given the incoming beams and energies, depend only the gluino mass and the strong coupling constant. This process would be dominating for most of the masses considered here (m$_{\tilde{\rm g}}<$ 2 TeV/$c^2$). A large number of gluinos (roughly $10^8$ for \mglu=100 \gevcc, $10^3$ for \mglu=1 \tevcc) is expected in one year of LHC running at low luminosity. They would be produced in a back-to-back configuration in the transverse plane, and their $p_T$ values are typically of the order of their own mass, i.e.~they may be relativistic but their mass is still far from negligible. Apart from the fraction of \glu g states, which is a free parameter, there is no preference for an R-hadron to be charged or neutral. Figure~\ref{fig:rhadprod} shows relevant distributions at event generator level (PYTHIA).

To find R-hadrons at the LHC, it is crucial to understand their interactions in matter. Several models exist to describe their interactions~\cite{models,aafke}. Here, the description of Ref.~\cite{aafke} is applied to study interactions and detection of R-hadrons in ATLAS. More details on the analysis of R-hadrons in ATLAS can be found in Ref.~\cite{atlaspaper}.

In the following, we discuss R-hadron production at the LHC and R-hadron interactions in matter. Focusing on ATLAS, signatures of single R-hadrons are summarized, and trigger issues are addressed. A study of the discovery potential is presented, and we end with a conclusion. 

\section{R-hadron interactions in matter}When traversing matter, all  R-hadrons undergo repeated  nuclear interactions. In addition, charged R-hadrons undergo continuous ionization losses and repeated Coulomb scatterings. A heavy hadron can be seen as a non-interacting heavy colored state, accompanied by a colored hadronic cloud of light constituents, responsible for the interaction. The typical energy scale of a nuclear interactions is expected to be small ($\lesssim 2$ GeV). The total cross section of R--meson scattering off a nucleon is taken to be about 24 mb~\cite{aafke}, and that for R-baryons 36 mb. In a nuclear interaction, they may flip charge (e.g.  $\rm uud + \tilde{\rm g}u\bar{d}\rightarrow udd+ \tilde{\rm g}u\bar{u}$) and baryon number (e.g.  $\rm udd + \tilde{\rm g}d\bar{u}\rightarrow u\bar{\rm u}+ \tilde{\rm g}ddd$ ). The latter process is important, since it is kinematically favorable, hence, heavy mesons always convert into heavy baryons, but not the other way around. A detailed discussion of the model as well as its GEANT3 implementation of the model used can be found in Ref.~\cite{aafke}.

\section{Detection in the ATLAS detector}
To study signatures in the different ATLAS subdetectors, singly charged R-mesons (most redundant hadrons in the hadronization) of various masses (100~\gevcc\ to 1900~\gevcc) and $p_T$ values have been fully GEANT3 simulated and reconstructed. The results below are independent of the event topology and are valid for heavy hadrons arising from any kind of color octet.

\begin{figure}[tp]
\begin{picture}(0,110)
\put(20,-10){\epsfxsize45mm\epsfbox{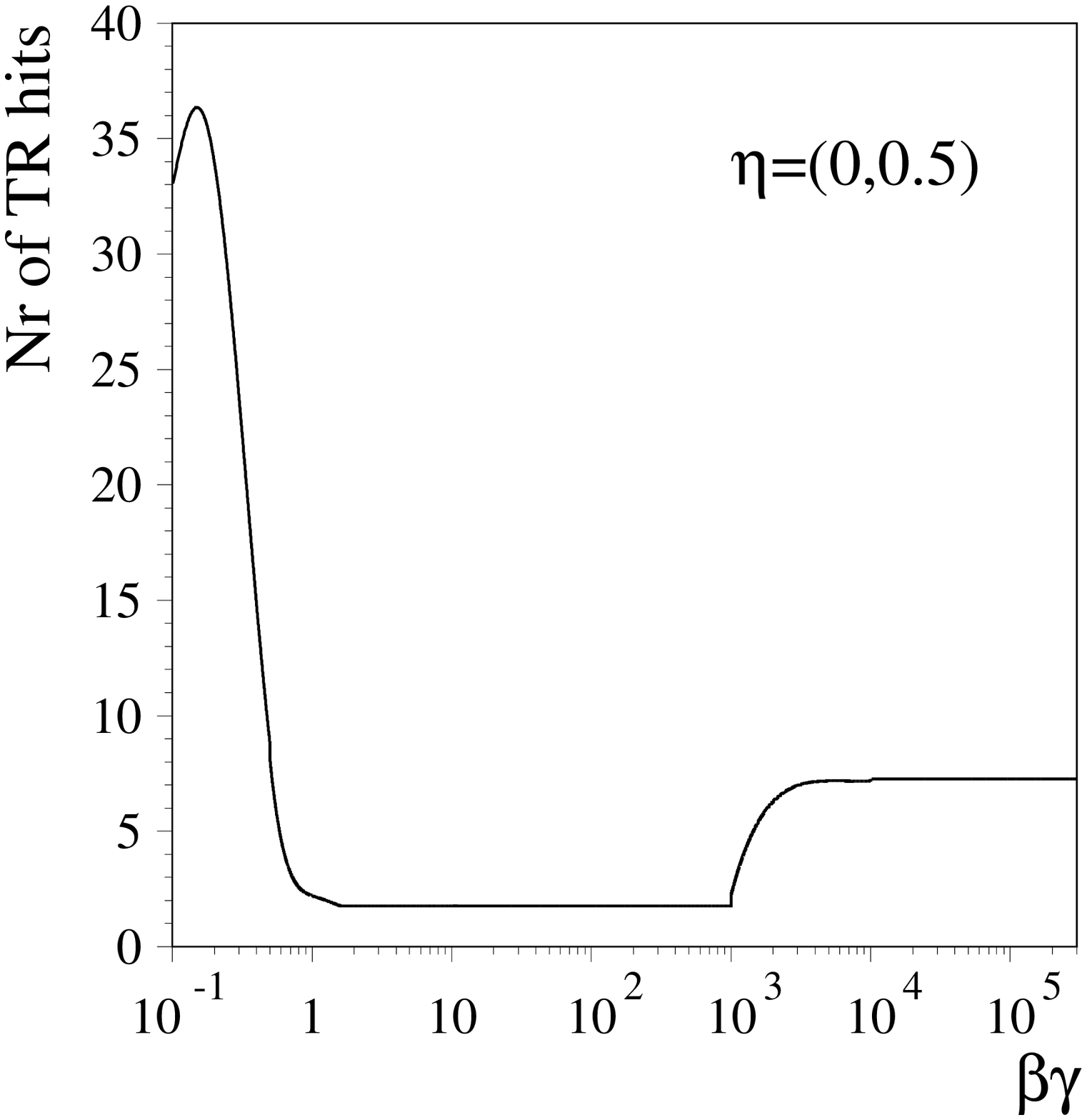}}
\put(170,-10){\epsfxsize45mm\epsfbox{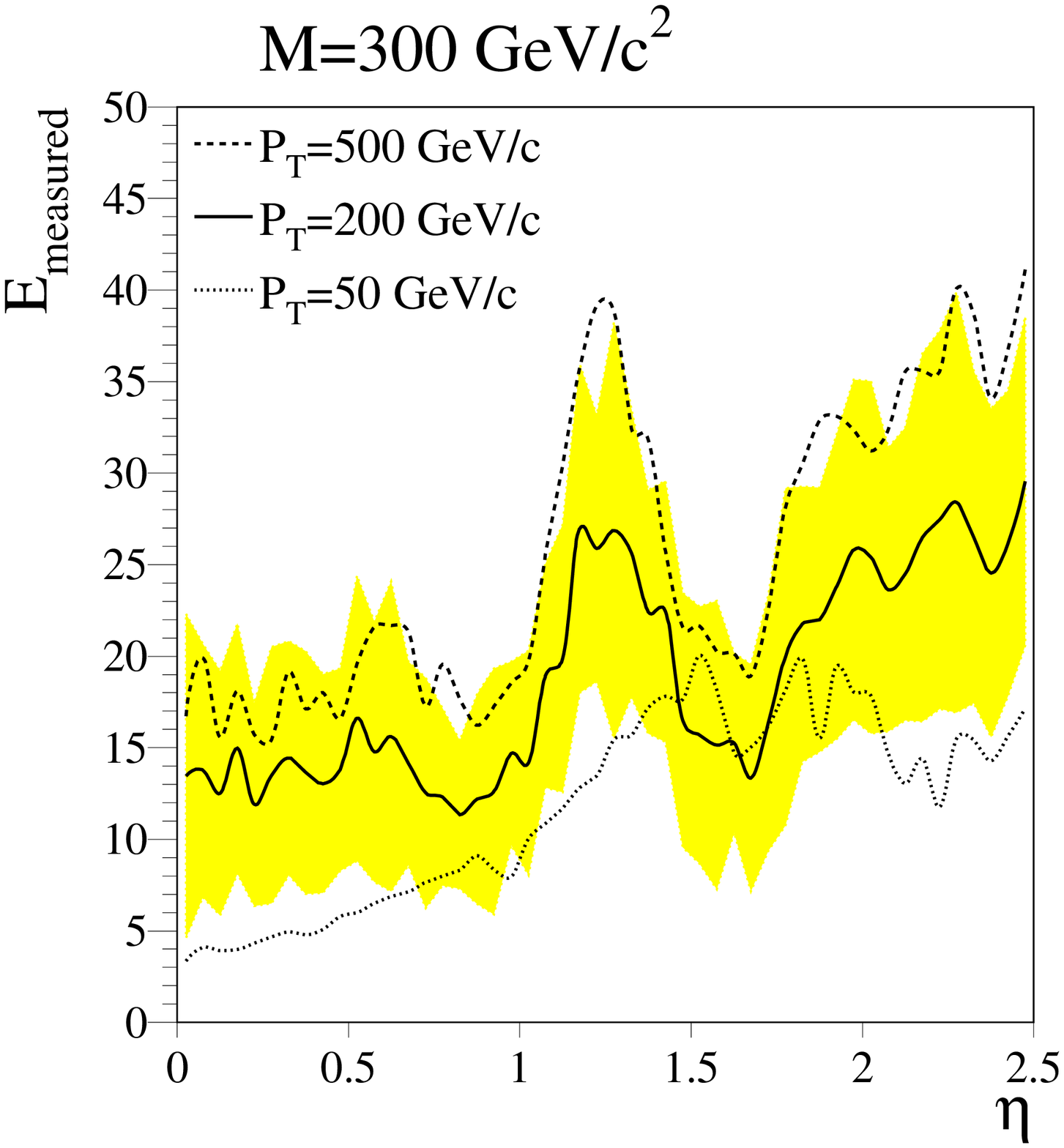}} 
\put(320,-10){\epsfxsize45mm\epsfbox{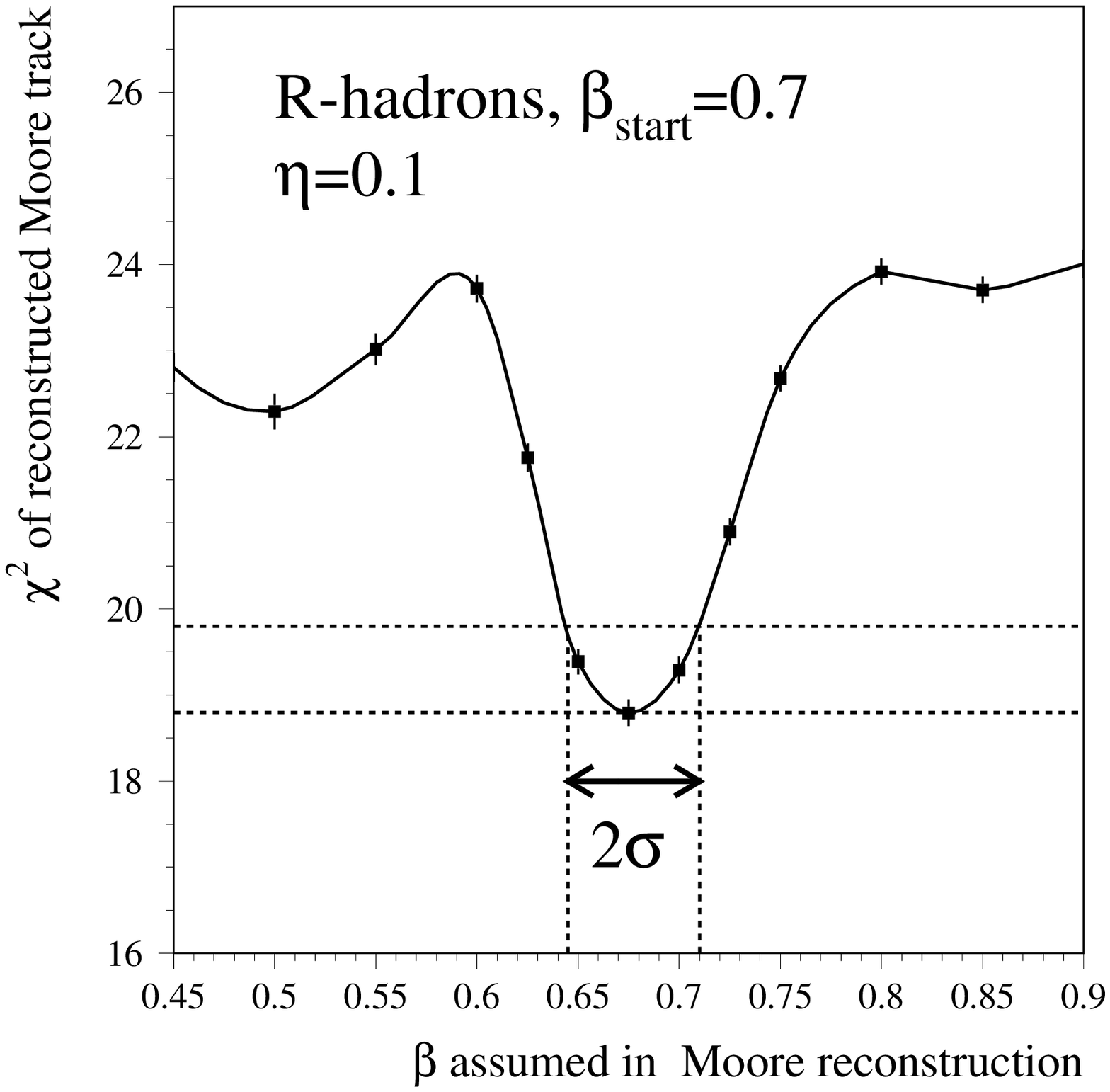}}
\end{picture}
\caption[ ]
{\protect\footnotesize
Left: average number of HT hits as function of $\beta\gamma$ of a particle 
   at central pseudo-rapidity. Center: energy loss for penetrating R-hadrons. The shaded area represents the spread in the energy deposit for R-hadrons with
  mass 300 \gevcc\ and $p_T$ of 200 \gevc. Right: $\chi^2$ from track reconstruction as function of $\beta$.
\label{fig:straws}}
\end{figure}

In the tracking system, R-hadrons would be characterized by their high $p_T$. The $p_T$ resulution of the ATLAS tracking system has been parameterized. Also, signatures of R-hadrons in the Transition Radiation Tracker (TRT) have been studied.
 A slow charged particle results in
large ionization energy deposits in the TRT. 
TRT hits being registered using two discriminator levels, allows R-hadron identification by counting the number of straw hits exceeding  the high threshold (HT) discriminator level (see Fig.~\ref{fig:straws}). 
 An other possibility is by measuring the time which a signal is above the low threshold discriminator level~\cite{aafke} (large for a heavily ionizing particle).
 
In the ATLAS calorimeters, R-hadrons undergo about 15 interactions (depending on $\eta$), whereby they may flip charge or baryon number. The total energy deposit measured in the ATLAS calorimeters is displayed as function of $\eta$ for various R-hadrons masses and $p_T$ in Fig.~\ref{fig:straws}. The energy deposits remain fairly constant along the trajectory of the particle (if indeed $\beta$ is not too small so that nuclear rather than electromagnetic losses dominate) and is much smaller than the momentum, in other words the E/p ratio is small, contrary to that of pions and electrons. A small fraction of R-hadrons is stopped inside the calorimeter.

The most crucial subdetector is the muon system. Time-of-flight measurements are possible thanks to the staggered arrangement of the MDT's (Monitoring Drift Tubes) and their good timing precision ($\simeq$ 1 ns). As can be seen in Fig.~\ref{fig:straws}, the $\chi^2$ of the fitted trajectory is sensitive to the particle velocity, thus providing a tool to reconstruct $\beta=v/c$~\cite{giacomop}. The resulting resolution of the R-hadron velocity in the production vertex is found to be $\sigma(\beta)/\beta^2=0.067\pm0.029$.

\subsection{Trigger issues}
Charged heavy particles reaching the muon system can trigger the muon trigger~\cite{nisati}. The efficiency of this trigger is studied by requiring a (charged) R-hadron not only to fire the trigger within the temporal gate of 18 ns~\cite{nisati}, but also to be recorded in the correct event, i.e. the R-hadrons delay compared to that of a muon must be less than 25 ns. This corresponds to a requirement of $\beta>0.5$. Efficiencies depend on the R-hadron mass, but are roughly of the order of 30\% and 60 \% for triggering on one and two charged R-hadrons, respectively. An extra complication is that R-hadrons undergo nuclear reactions in the muon system due to the support structures~\cite{atlaspaper}.

Also, calorimeter triggers are crucial for R-hadrons detection. Although the R-hadrons are produced roughly back-to-back, an imbalance would arise if one charged (resulting in a high $p_T$ track) and one neutral R-hadron (only depositing energy) would be detected. 

For very slow particles, effects of the time delays in the detector are complicated. The condition $\beta>0.5$ allows for a adequate reconstruction of the signal in all subdetectors. 
\section{Discovery potential}
\begin{figure}[t!]
\vspace{-0.3cm}
\begin{center}\epsfig{file=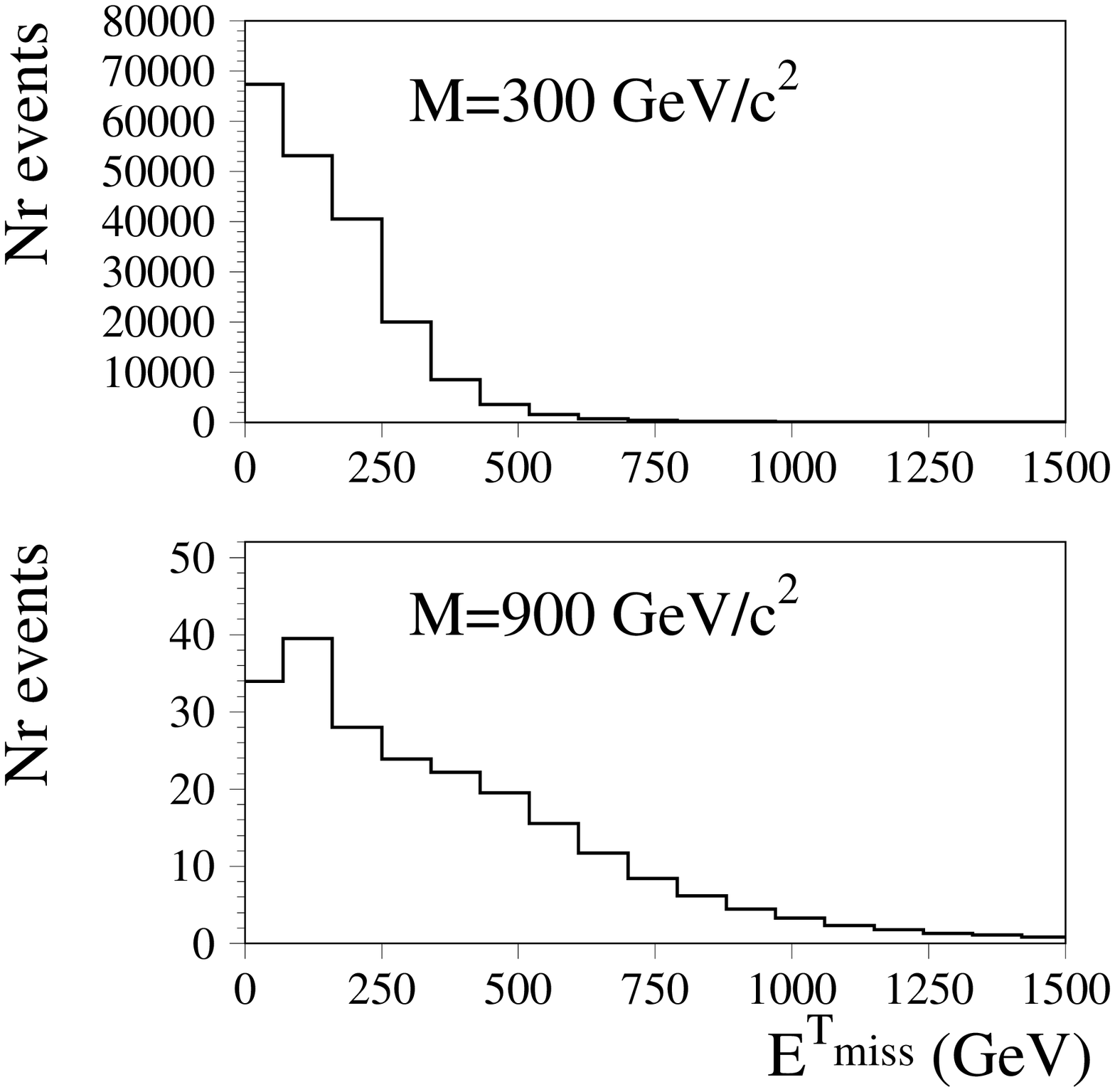,height=4.6cm,width=5cm}\epsfig{file=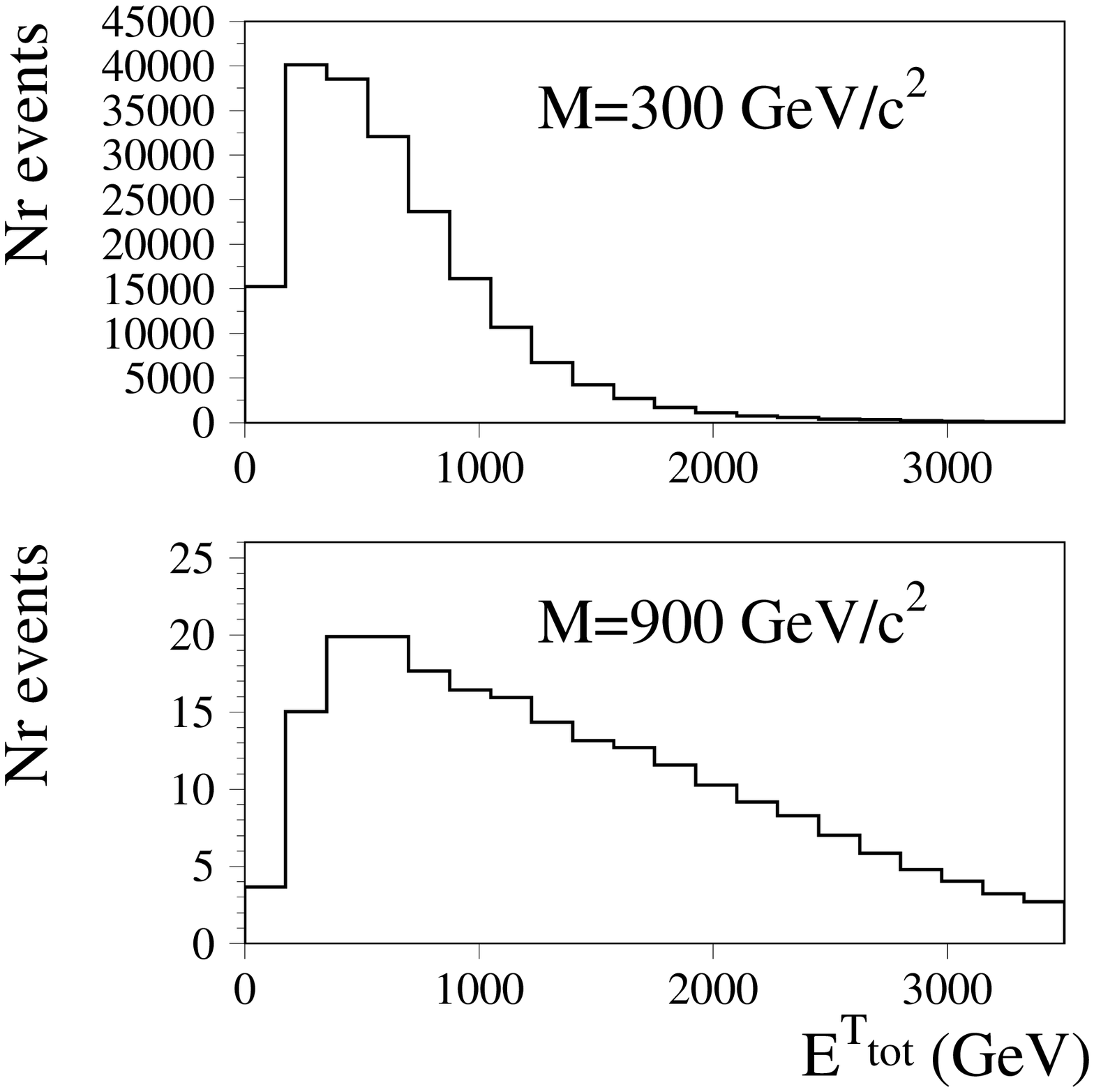,height=4.6cm,width=5cm}\epsfig{file=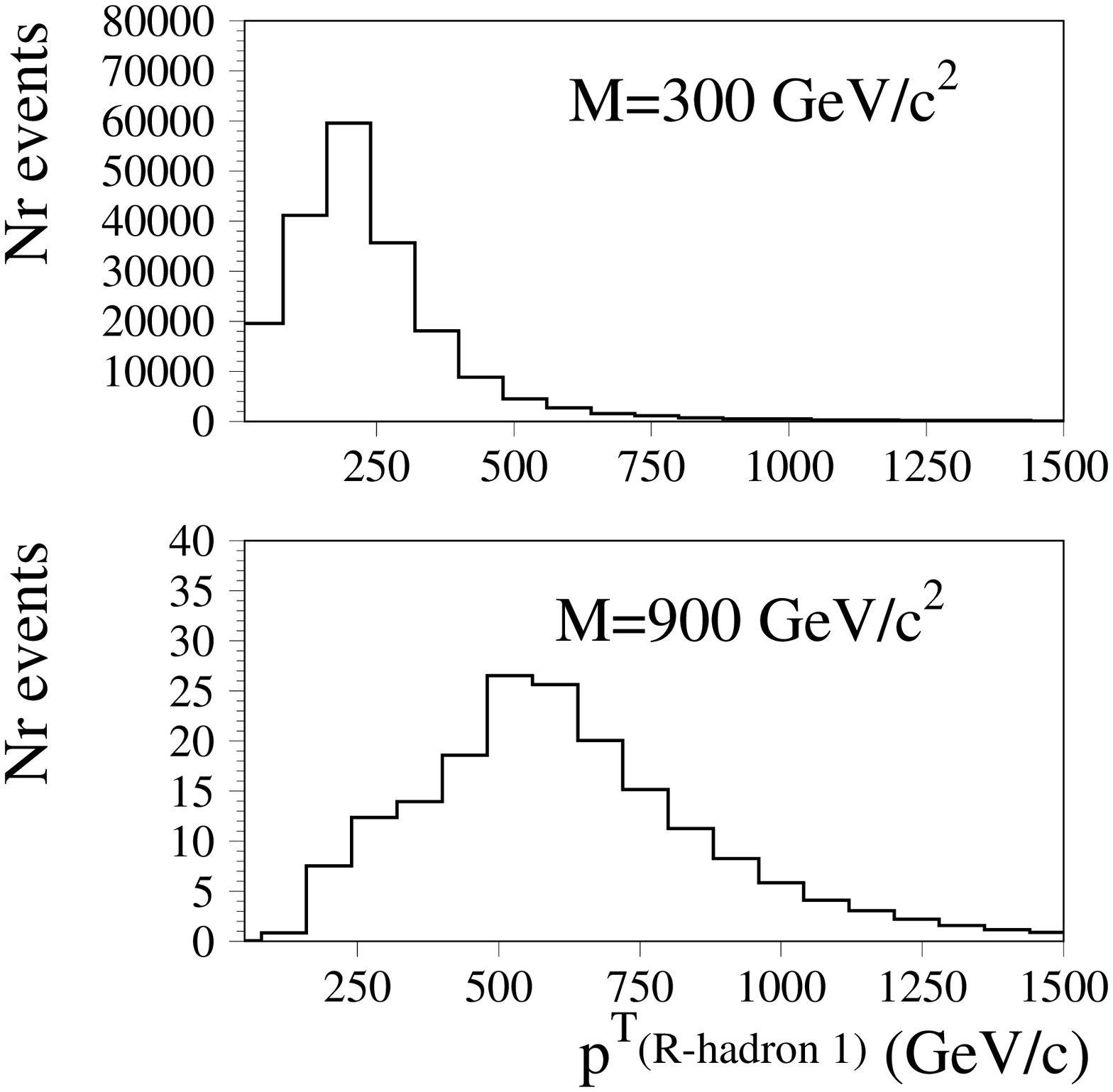,height=4.6cm,width=5cm}\\
\end{center}
\vspace*{-0.5cm}
\caption[]
{\protect \footnotesize  The missing transverse energy, the total visible energy, and the transverse momentum of the R-hadron or muon in the muon chambers, after high level trigger requirements for signal events. The number of events in these plots corresponds to an integrated luminosity of 1~fb$^{-1}$. \label{fig:11}}
\end{figure}

No official parameter space points exist. We concentrate on ~\chancc, and investigated masses between 100 \gevcc~and 2 TeV$/c^2$, for LHC running at low luminoscity (3 years, 10 $\rm pb^{-1}$/y). Fast simulation, based on parameterizations from fully simulated single R-hadrons, is used to simulate and perform analysis. Prior to selection, events are subjected to a 'psuedotrigger' implementing the ATLAS trigger menues relevant for R-hadrons. The analysis is performed on the basis of \emph{event topology} alone, not requiring a fine simulation of the response of the ATLAS detector. Background processes considered were QCD, W, Z and WW/WZ/ZZ. Cuts on the $p_T$ (Fig.~\ref{fig:11} left) of a particle as recorded by the muon chambers, the missing transverse energy (Fig.~\ref{fig:11} center), and the total visible energy of the event (Fig.~\ref{fig:11} right) allow R-hadrons to be discovered ($S/\sqrt{B}>5$) up to 1400 GeV. Low mass R-hadrons, thanks to their large production cross sections, could already be discovered after only a few days of data taking.

When using more involved information, such as requiring the R-hadrons delay in the muon stations to be more than 3 ns than that of a particle with $\beta=1$, R-hadrons up to 1700 GeV could be discovered. Including contributions from $q\bar{q}\rightarrow \tilde{g}\tilde{g}$, the mass reach would be extended to 1600 using global event variables and 1830 \gevcc\, including TOF information. Among uncertainties influencing the discovery limit are the probability to obtain a $\tilde{g}g$ state during hadronization, the handling of events with neutral R-hadrons in the inner detector and charged
R-hadrons in the muon system by the event reconstruction, and the R-hadron nuclear interactions inside the muon system. However, these would only lower the discovery potential by about 100 GeV.

\section{Conclusion}
This  study has addressed many aspects important for the detection of R-hadrons. If R-hadrons exist with masses below 1800 GeV, they will be discovered at early stages of LHC running. 
\section{Acknowledgements}
I would like to thank J\o rgen Beck Hansen for his important contribution to these results, and for suggestions to this note.

\end{document}